\documentclass{article}

\pdfoutput=1
\usepackage{geometry}
\usepackage[pdftex]{graphicx}
\usepackage[utf8]{inputenc} 
\usepackage[T1]{fontenc}    
\usepackage{url}            
\usepackage{booktabs}       
\usepackage{amsfonts}       
\usepackage{bm}

\usepackage{nicefrac}       
\usepackage{microtype}
 \usepackage{algorithm}
\usepackage{algpseudocode}                                      

\usepackage{color}
\usepackage[compact]{titlesec}
\usepackage{wrapfig,lipsum,booktabs}
\usepackage{lipsum}
\usepackage{sidecap}
\usepackage[draft]{fixme}

\usepackage{caption}
\newif\ifcomments
\commentstrue
\ifcomments
\newcommand{\pjm}[1]{{\color{blue}{#1}}}
\else
\newcommand{\pjm}[1]{}
\fi

\ifcomments
\newcommand{\mn}[1]{{\color{red}{#1}}}
\else
\newcommand{\mn}[1]{}
\fi

\ifcomments
\newcommand{\mnl}[1]{{\color{red}{($\leftarrow$ #1)}}}
\else
\newcommand{\mnl}[1]{}
\fi

\graphicspath{{Figures/}{./}}

\title{Compressing Networks with Super Nodes}

\author{Natalie Stanley$^1$ \and Roland Kwitt$^2$ \and Marc Niethammer$^1$ \and Peter J. Mucha$^1$}
\date{%
    $^1$University of North Carolina, Chapel Hill\\%
    $^2$University of Salzburg\\[2ex]%
    \today
}

\begin{document}

\maketitle

\begin{abstract}
 Community detection is a commonly used technique for identifying groups in a network based on similarities in connectivity patterns. To facilitate community detection in large networks, we recast the network to be partitioned into a smaller network of `super nodes', each super node comprising one or more nodes in the original network. To define the seeds of our super nodes, we apply the `CoreHD' ranking from dismantling and decycling. We test our approach through the analysis of two common methods for community detection: modularity maximization with the Louvain algorithm and maximum likelihood optimization for fitting a stochastic block model. Our results highlight that applying community detection to the compressed network of super nodes is significantly faster while successfully producing partitions that are more aligned with the local network connectivity, more stable across multiple (stochastic) runs within and between community detection algorithms, and overlap well with the results obtained using the full network.
  \end{abstract}

\section{Introduction}
Networks appear across disciplines as natural data structures for modeling relations between entities, such as regulatory interactions between genes and proteins, and social connections between people. In practice, most networks are large and have intricate substructures that can be interrogated for further insights about the underlying data. A common practice for extracting subgraphs of interest is community detection \cite{fortunato2010,fortunato,portercommunity}, which aims to partition network nodes into groups based on the group-level connectivity patterns. For example, assortative communities are typically defined through some measure of a greater weight of within-group edges compared to those between groups. Communities can be identified through various optimization problems, including likelihood maximization \cite{sbmOrig,bigclam,tiagosbm} and quality function optimization \cite{louvain,newmanmodularity,gleich}, or from spectral properties of associated matrices \cite{spectralCommunity}.

In this paper, we specifically examine how the compressed version of a network can be handled by modularity maximization and likelihood maximization. 
To maximize modularity in a network with $N$ nodes, one seeks to find the node-to-community assignment vector, ${\bf z}=[z_{1},z_{2},\dots z_{N}]$ that maximizes modularity, $Q$, defined by
\begin{equation}
Q=\frac{1}{2M}\sum_{i,j}[a_{ij}-\gamma \frac{k_{i}k_{j}}{2M}]\delta(z_{i},z_{j}),
\end{equation}
where $a_{ij}$ encodes presence and weight of a possible edge between $i$ and $j$, $k_{i}=\sum_{j}a_{ij}$ gives the strength of node $i$, $\delta(z_{i},z_{j})=1$ if nodes $i$ and $j$ have the same community assignment and 0 otherwise, $M$ denotes the number of edges and $\gamma$ is the resolution parameter controlling community sizes.
Alternatively, to find community structure maximizing likelihood, one can fit a stochastic block model (SBM), which assumes that connectivity patterns of $N$ nodes in a network with $K$ communities can be modeled according to their community memberships through a probability matrix ${\bm{\pi} }$, where ${\pi_{kl}}$ gives the edge connection probability between communities $k$ and $l$. Assuming no corrections due to node degrees (degree-corrected versions also exist), one seeks to $(\bf z)$ maximize,
\begin{equation}
p({\bf a}\mid {\bf z}, {\bm \pi})=\prod_{i\ne j}^{N}\prod_{k,l}^{K}({\pi}_{kl}^{a_{ij}}(1-\pi_{kl}^{1-a_{ij}}))^{\delta({z_{i}=k,z_{j}=l})},
\end{equation}

where $\delta({z_{i}=k,z_{j}=l})$ is 1 if $z_{i}=k$ and $z_{j}=l$, and 0 otherwise. One can maximize this objective with the expectation maximization (EM) algorithm, belief propagation, and efficiently with an iterative approach which, similar to Louvain, agglomerates blocks of nodes at each iteration \cite{tiagosbm}. Specifically, at each iteration a Metropolis-Hastings accept-reject sampling approach is used to compute a probability of merging blocks based on how the merge affects the likelihood. These agglomerative heuristics simplify a computationally challenging task but can still be time consuming for large networks and often give rise to large variability in returned partitions across multiple runs of the algorithms. We seek to explore how a compressed network representation can improve some of these issues. 

The identification of communities in networks is in some ways similar to multi-label image segmentation, which aims to partition a grid of pixels into contiguous regions corresponding to objects in the image. In this sense, each segmented region can be viewed as a community \cite{browet}. To speed up segmentation for large images, a popular approach is to avoid computing segmentations at the pixel level and instead reformulate the segmentation problem based on larger-scale image primitives that are likely part of the same partition. Specifically, this can be accomplished by {\it super pixels} that aggregate pixels together in a way that faithfully adheres to image boundaries, maintaining or improving segmentation accuracy \cite{slic}. The SLIC super pixel method \cite{slic} chooses seed pixels across the image's pixel grid to serve as the super pixel centers and then iteratively grows out and recomputes based on aggregation with neighboring pixels with similar visual features. Motivated by super pixels, we wish to define seed nodes in networks that can be used as a starting point to grow out `super nodes' to define a new, smaller network upon which we apply standard community detection algorithms. Creating a direct analog of super pixels in networks is challenging because the inherent geometry of a network can be quite different from the grid layout of an image (where simple neighborhood structures such as 4- or 8-neighborhoods are typically used), and we need ensure seeds are well distributed across the network. Further, while super pixels are largely constrained by the structure of the pixel grid (i.e. proximity between pixel pairs matter), their definition also incorporates extra image features to refine members of a super pixel set, whereas in network community detection we typically only have the edges of the network to work with. Finally, the performance of a super pixel representation of an image can be objectively validated from the quality of the corresponding segmentation result, with reference to human-specified objects in images; in contrast, community detection is typically an unsupervised, exploratory data analysis technique with limited available notions of `ground truth' \cite{jureTruth}. As such, we must develop measures that can be used to validate the quality of the super node representation.

Our approach in defining a smaller network of super nodes is a form of network compression. A useful approach developed previously is the `KeepAll' method \cite{gilbert}, which seeks to prioritize a set of nodes and retain only the smallest set of additional nodes required for the induced subgraph of prioritized nodes to be connected. Two other network compression methods related to our objective in defining super nodes are explored in Refs.~\cite{supergenomic,SuperNodeSide}; but these approaches differ from our proposal in that they seek to define super nodes along with additional side information about relationships between node pairs. Ref.~\cite{supergenomic} describes `super genomic network compression' to reduce the number of edges in a large protein interaction network. To do this, the authors identify `clusters of orthologous  groups' of proteins, or proteins that give rise to similar functions in different species and originated from a common ancestor. Members of an orthologous group are connected as a star network, with the center node as one member of the orthologous group. Furthermore, edges between orthologous groups are replaced by a single weighted link reflecting the pairwise group evolutionary similarity. Ref.~\cite{SuperNodeSide} defines super nodes by defining `must link' and `cannot link' constraints between pairs of nodes, agglomerating as many nodes as possible sharing must link constraints while being cautious about agglomerating nodes that cannot link. 

\begin{figure}[t!]
\centering
\includegraphics[scale=0.4]{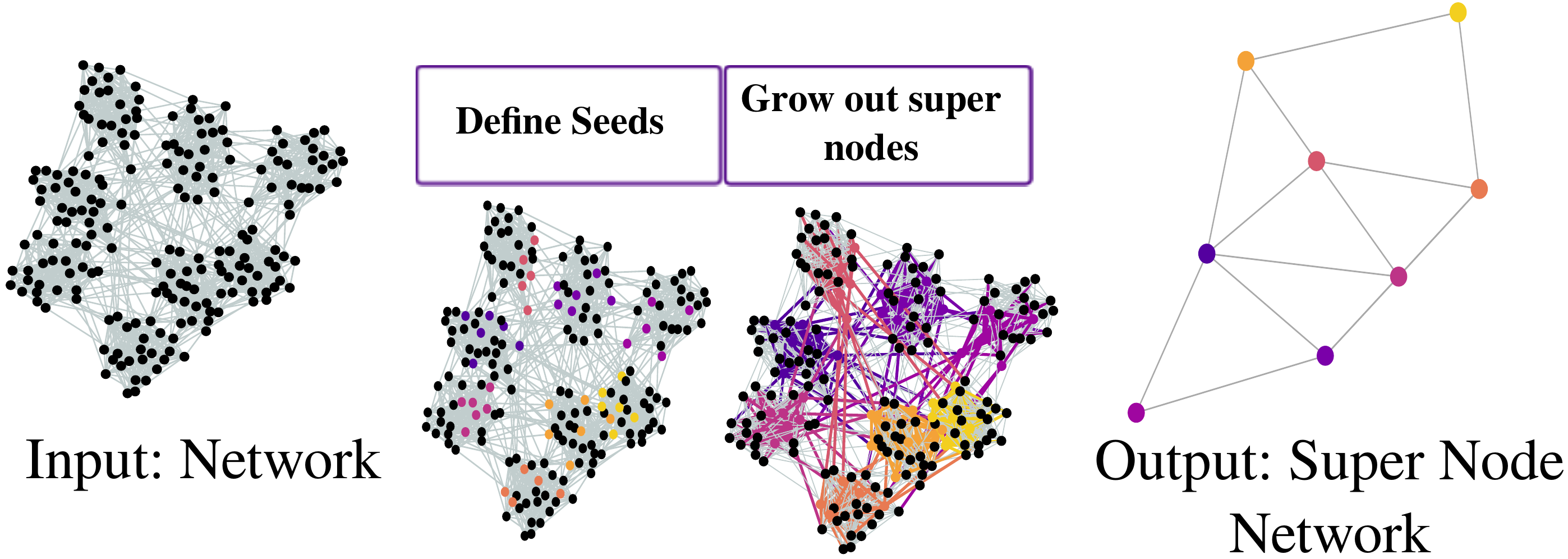}
\caption{{\bf Defining super nodes}. To define the super node representation of a network, we select $S$ seeds and agglomerate local regions around them to create super nodes. This then leads to a new network with weighted edges between the $S$ super nodes upon which community detection can be more efficiently applied.}
\label{Fig1}
\end{figure}

To define the super node representation of an $N$-node network, we first select $S\ll N$ seed nodes, agglomerate the remaining $N-S$ nodes with the seeds to create super nodes, and specify the network between these super nodes. Community detection can then be applied to the $S$-node network representation. Figure \ref{Fig1} visualizes this approach, with details provided in Sec.~\ref{sec:method}.

The aims of this work are twofold: \emph{First}, we seek an effective way to define a super node representation of a network that can then be used in standard community detection algorithms. \emph{Second}, we wish to establish a principled approach to validate the quality of a super node representation. Our approach to defining super nodes aims to accomplish the following.
\begin{enumerate}
\setlength\itemsep{0mm}
\item {\bf Decreased runtime for community detection}: Even though recently developed heuristics for maximizing modularity \cite{louvain} and fitting SBMs \cite{tiagosbm} are highly efficient relative to previous approaches for performing the same computational optimizations, these methods can still be time consuming for large networks. We aim to  reduce runtime for large networks, moving most of the computational cost in practice from tasks scaling with the size of the network to alternatives scaling with the (much smaller) size of the super node representation.
\item {\bf Decreased stochastic variability of community detection algorithm output}: In large networks, there is often significant variability across multiple runs of the same algorithm (employing computational heuristics to solve NP-Complete optimizations), as well as differences between various community detection algorithms. We expect applying community detection to a well-chosen super node representation to decrease the observed variability.
\item {\bf High local agreement}: In defining super nodes, we inherently assume that the identified communities should agree with the local network connectivity in that members of a neighborhood should be more likely to have the same community assignment, provided that the super nodes were constructed to minimize aggregation across community boundaries.
\item {\bf Consistent with communities found using the full network:} The identified community structure should be relatively similar to the distributions of results that would have been obtained through applying community detection to the full network.
\end{enumerate}

\textbf{Organization}.
In section~\ref{sec:method}, we describe the specific algorithms we employ for choosing seeds, growing these out to super nodes, and specifying the network of super nodes. Our method of selecting seed nodes is a novel application of the CoreHD algorithm \cite{CoreHD}, previously used for network decycling and dismantling. Section~\ref{sec:results} presents experiments used to understand and quantify the quality and usefulness of the super node representation, addressing the four objectives listed above. 
Finally, in section~\ref{sec:conclusion}, we discuss other possible applications of super nodes and directions for future work.

\section{Method for Generating a Super Node Representation of a Network}
\label{sec:method}
We create a super node representation through three steps, outlined in Figure \ref{Fig1}. First, we define seeds. Next, we `grow' super nodes by assigning the remaining nodes to seeds. Finally, the network of super nodes is defined by agglomerating edges, and can then be used in a community detection algorithm. 
\subsection{Identify Seeds in the Network}
To define $S$ seeds, we aim to identify a set of nodes $S^{*}$ that are individually central to the network and to their communities, and that are well separated from one another. Such problems are related to influence maximization, where one identifies a small number of nodes in the network from which to effectively spread influence or diffuse messages across the network \cite{jungirie,kempe}. The most naive approach is to select nodes with highest degree, and this might be perfectly reasonable under various circumstances. Importantly, the selection of nodes with highest degree is computationally fast, requiring $O(M)$ operations, summing over the $M$ edges to calculate the degrees of the $N$ nodes. 

\begin{SCfigure}[2]
\centering
\includegraphics[width=0.25\textwidth]{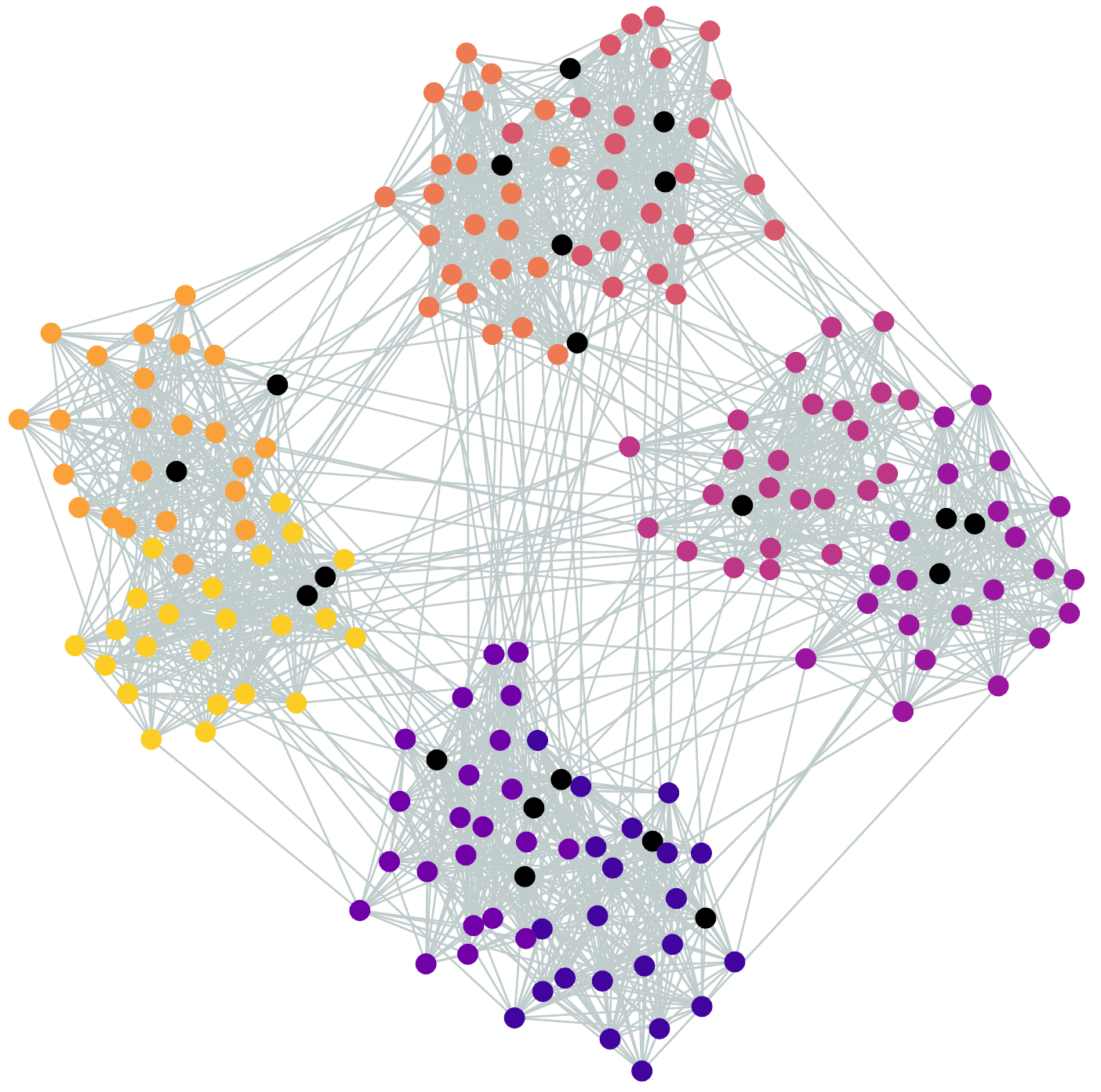}
\caption{{\bf Choosing seeds in a synthetic network}. The identification of 20 seeds with the CoreHD algorithm in a network generated from a stochastic block model with 8 communities. Seeds (black nodes) are well distributed across communities.}
\label{Synth}
\end{SCfigure}

At slightly higher computational cost, we employ the CoreHD algorithm, which nearly optimally identifies nodes in network decycling and dismantling \cite{CoreHD}. CoreHD recursively identifies the highest degree node in the 2-core (the maximal connected subgraph in which all nodes have degree at least 2 within the subgraph, obtained efficiently by recursively pruning away all nodes of degree less than 2). At each iteration, the removed node is added to our seed set and the now smaller 2-core and their degrees are recomputed. The difference between selecting highest degree nodes and CoreHD for our present task may be small, both in terms of result and computational cost. In particular, because we will only select $S\ll N$ seeds, there is reduced opportunity for the removals to lead to subgraphs with substantial differences between degree order in the graph and its 2-core [cf.\ selection of $O(N)$ removals in network dismantling]. Indeed, in our experience, simply selecting the highest degree nodes as the seeds often works well in practice. Because of the minimal extra computational cost for computing the 2-core, we use CoreHD for all of our results shown here. In Figure \ref{Synth}, we plot a sample network generated from the stochastic block model, with nodes colored according to community assignments and black nodes indicating seeds, showing the seeds are distributed across all 8 communities. In supplementary figures 2 and 3, we compare the communities obtained selecting seeds by different methods with the full network communities.

\subsection{Grow Super Nodes Around Seeds}
Once the set of seeds, $S^{*}$, is defined, we `grow' them out  agglomerating nearby nodes to build the super nodes. We formally define a super node as a subset of one or more nodes from the original network, ${\mathcal X}$. To do this, seeds are grown out to engulf nodes in increasing neighborhood orders until either all nodes are assigned to a super node center or until a user-defined number of neighborhood orders has been considered. The maximum order, $o_{max}$, can be specified to control the maximum order neighborhood to consider in building the super nodes. If after $o_{max}$, there are still unassigned nodes, the unassigned nodes are not used to build the new super node network and are all ultimately assigned to the same periphery community as they are not considered relevant to the network core.  Depending on the number of chosen super nodes, $S$, the degree distribution of the original network and the quality of the chosen seeds at collapsing the network, different networks will require repeating the agglomeration process for different neighborhood orders if one wants to ensure every node is assigned to a super node.  
The output is a vector, ${\bf s}$ of length $N$, which gives the node-to-supernode assignments for the nodes in the original network. 

\subsection{Create Network of Super Nodes}
Finally, after growing the super nodes, we must create a new network representation of the super nodes. To do this, we create a weighted network, $\mathcal{W}$, where each super node is a node and the weight of the edge between a pair of distinct super nodes is the total weight of edges in the original network ${\mathcal X}$ between pairs of nodes assigned to the respective  super nodes. For pairs of super nodes whose members have no edges between them in ${\mathcal X}$, there is no corresponding edge in ${\mathcal W}$. By definition, we construct ${\mathcal W}$ with no self loops.

After applying community detection to the super nodes, their community assignments are mapped back to their constituent $N$ nodes of the original network, ${\mathcal X}$. In section~\ref{sec:results}, we use this final $N$-length matched community assignment, which we denote as ${\bf z}$. We consider the node-to-community assignment ${\bf z}^{Full}$ obtained by applying community detection to the full network and the mapped result ${\bf z}^{SN}$ obtained by applying community detection to the super node representation.

\section{Results}
\label{sec:results}
To demonstrate the effectiveness of super nodes, we performed experiments to analyze the runtime, partition variability, alignment of communities with local network connectivity, and agreement with the communities obtained from the full network. We considered 9 network data sets (see Table~\ref{table:dataset})  from the Stanford Network Analysis Project database\footnote{\url{https://snap.stanford.edu/data}} (Enron, Amazon, Dblp, Email, BrightKite, 
\begin{wraptable}{r}{.5\textwidth}
\begin{center}
\scalebox{0.7}{
\begin{tabular}{ |c|c|c|c| } 
\hline
\textbf{Dataset} (* indicates subgraphs) & \# Nodes & \# Edges \\
\hline
As22* (Internet)  & 22,801 & 48,270 \\ 
Enron* & 32,374 & 178,195 \\ 
CMatter* (Condensed matter 2003 collab.)& 17,816 & 83,337\\ 
Dblp* (com-DBLP) & 150,801 & 639,330\\
Amazon* (com-Amazon) & 77,463 & 209,887\\
Email (email-EuAll) & 265,214 & 420,045\\
Stanford (web-Stanford) & 281,903 & 2,312,497 \\
Notre Dame (web-Notre Dame) & 325,729 & 1,497,134\\
BrightKite (loc-BrightKite) & 58,228 & 214,078 \\
\hline
\end{tabular}
}
\end{center}
\caption{Network data characteristics.\label{table:dataset}}
\end{wraptable}
Stanford, Notre Dame) and Newman's collection\footnote{\url{http://www-personal.umich.edu/~mejn/netdata/}} (As22, CMatter). 
We treat all networks as undirected. To explore a range of network sizes, in some of these cases we used large subgraphs defined by the union of all nodes of degree $\geq 2$, their neighbors, and next nearest neighbors. We use the Louvain algorithm\footnote{as implemented in \url{https://github.com/vtraag/louvain-igraph}} for modularity maximization \cite{louvain} and the stochastic block model (SBM) inference\footnote{as implemented in \url{https://graph-tool.skewed.de/}} described in \cite{tiagosbm}. In all experiments, we created the super node representation with $S=500$ super nodes. 

\subsection{Objectively Comparing Partitions on Possibly Different Scales}
A challenge in directly comparing the community partitions on the full and super node network representations is the difference in scales between the partitions. For example, using the full network typically produces significantly more communities than under the super node representation. In an attempt to compare community partitions with similar size distributions in the subsequent experiments, we can adapt the scales obtained from the Louvain algorithm and SBM fitting.

In community detection with the Louvain algorithm, we identified comparable resolution parameters (controlling community size) to apply to the full network that would produce a size distribution agreeing as much as possible with the community partition in the super node network. Further, we compute all following experiments using both the default resolution parameter and the `matched' parameter. While the default resolution parameter is $\gamma=1$, in our analyses we computed partitions of the full network using several different $\gamma\in[0.05, 2.5]$. To choose the matched resolution parameter on the full network, we first find the community partition using the super node representation. For each partition, we then rank nodes based on the sizes of the communities to which they belong. For each partition of the full network (at different resolution parameters), we then consider the similarity of this ranking with that from the super node communities, measuring this similarity by Kendall's tau correlation. We identify the resolution parameter producing the highest Kendall's tau correlation, referring to this resolution parameter as the `matched parameter' in the remainder of the text, while we refer to the standard $\gamma=1$ as the `default' resolution parameter.

In fitting SBMs, we chose to fit a model with the same number of blocks that was found in the super node representation using the standard optimization and model selection strategies discussed in \cite{tiagosbm}. We refer to the `matched' version as that using the number of blocks identified by the model selection on the super node representation, while the `default' result is obtained using the model selection strategy on the full network. In subsequent experiments, we compare both the `matched' and `default' versions to ensure our results are not artificially influenced by the scale of the community sizes.

\subsection{Experiments}
\begin{figure}
\centering
\includegraphics[scale=0.4]{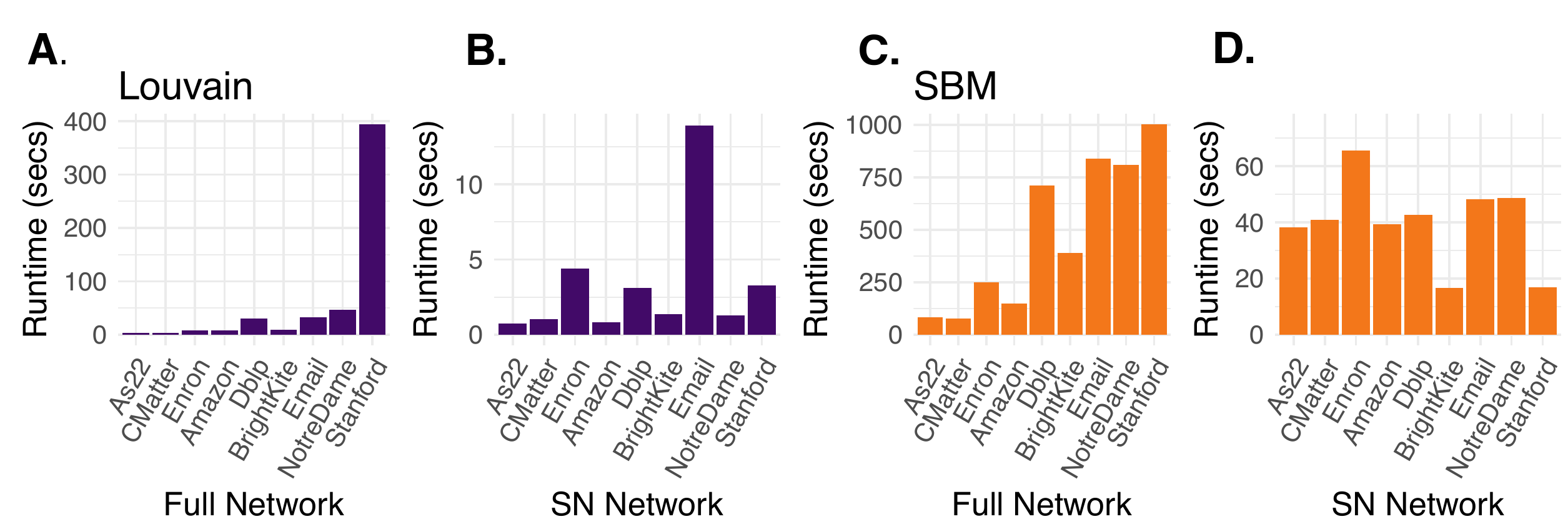}
\caption{{\bf Runtimes}. We compare community detection runtimes (in seconds) with the Louvain algorithm and by fitting an SBM on the full networks and super node representations for the 9 data sets. {\bf (A)} Louvain on the full network. {\bf (B)} Louvain  on the super nodes. {\bf (C)} SBM on the full network. {\bf (D)} SBM on the super nodes.}\label{RT}
\end{figure}

For each of the 9 networks, we recorded the runtime required to identify communities with the Louvain algorithm and stochastic block model inference procedure under the full and super node representations (Figure \ref{RT}). The Louvain algorithm is fast and scales well, at $O(M)$ per iteration for $M$ edges, with its relative speed and high modularity values contributing to its popularity. While the reported runtimes may seem quite modest, in practice it is common to run many realizations of the algorithm (hundreds, thousands, or even more for large networks) to explore resolution parameters and stochastic variation due to node order in the heuristic. We note a large increase in runtime for the full Stanford network, with over 2 million edges. As also observed in the figure, fitting a stochastic block model, at $O(N{\ln}^{2}N)$ for sparse networks in this implementation \cite{tiagosbm}, becomes significantly slower on the full networks with more than 200,000 edges. 

While we see a significant improvement in community detection runtime from using super nodes for both methods, the benefit in the SBM fitting is particularly large, especially for the bigger networks (DBLP, Stanford, Email). In moving to the super node representation, we traded out large-coefficient scaling-with-$N$ community detection computations for those scaling with $S\ll N$ (with possible increases due to the increased density of the super node representation), at the cost of constructing the super node representation. In particular, we observe the SBM runtimes on super nodes appear to be relatively independent of $M$. We note that each of the three steps building our super-node representation is $O(M)$, so in the large graph limit the expected gain of our approach may be only a constant factor over Louvain iterations and the SBM fitting (up to logarithmic factors). In the present calculations, we have not endeavored to optimize the runtime to build our super node representations; even so, the three steps building the $S=500$ super node representation of the Stanford network in our current implementation together take $\sim 350$ sec with CoreHD and $\sim 200$ sec using highest degree nodes. While this alone might not seem like a large improvement compared to a single realization of running Louvain or fitting an SBM, the computational gains compared to generating multiple community partitions can be quite significant.  
 
\begin{figure}
\centering
\includegraphics[scale=0.4]{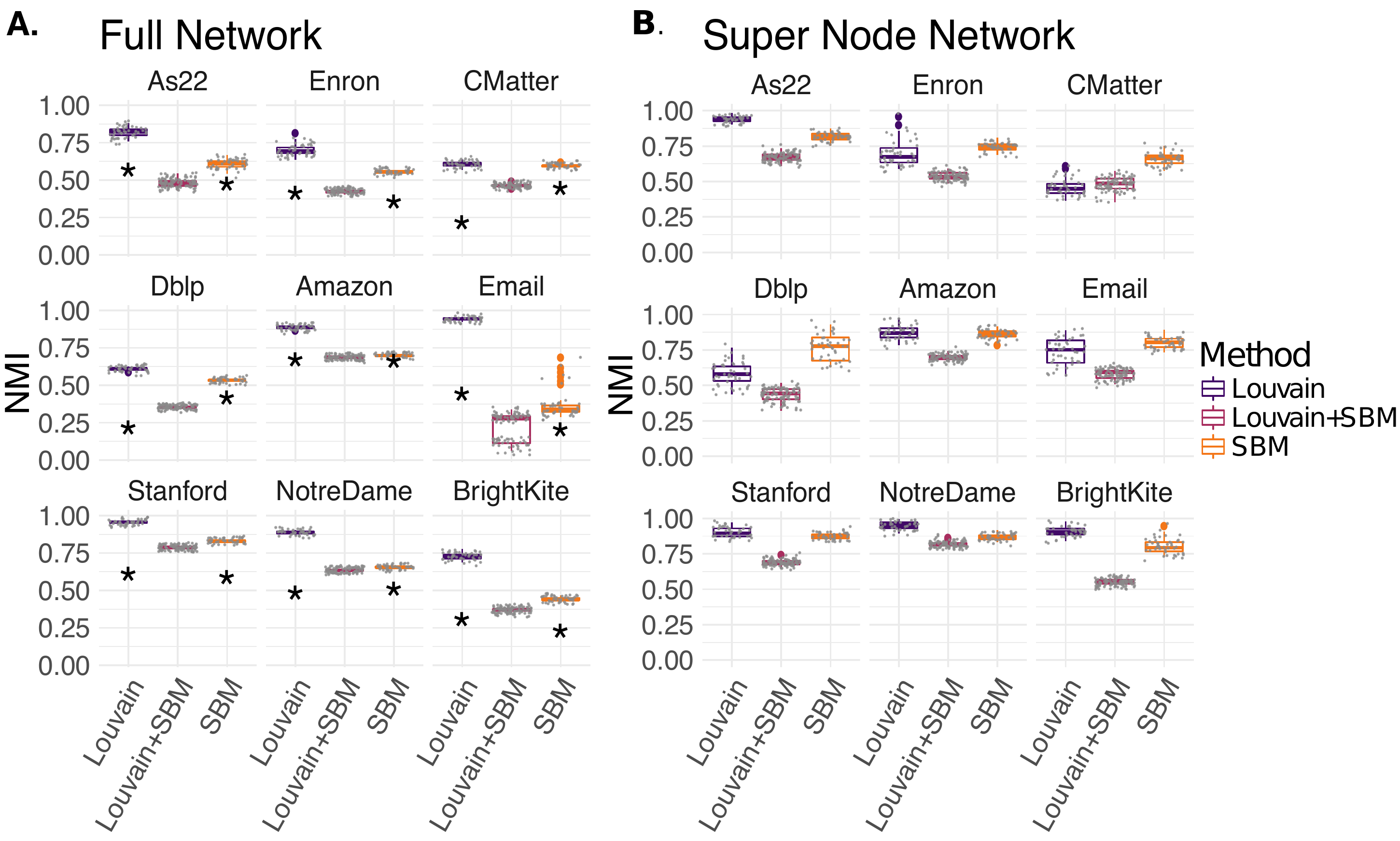}
\caption{{\bf Quantifying partition variability.} For each of the 9 networks, we obtained 10 different partitions using Louvain and fitting SBMs under default settings. We computed pairwise normalized mutual information (NMI) between partitions within and between algorithms. For each network, the purple, orange and maroon boxplots (and jittered points) show the pairwise NMIs between the 10 runs of Louvain, the 10 SBM fits, and each individual run of Louvain against the SBM fitting (Louvain + SBM). {\bf (A)} Within- and between-algorithm variability on the full networks. Stars in the Louvain and SBM column indicate the mean pairwise NMI computed between 5 super node partitions and a full network representation under Louvain and SBM, respectively [i.e., NMI$({\bf z}^{Full},{\bf z}^{SN})$ for multiple super node representations. {\bf (B)} Within- and between-algorithm variability on the super node representations. Overall, we observe decreased variability within and between runs of community detection with the super node representations compared to the full networks, while in the context of this general variability the super node results are comparable to the `default' communities from the full networks.}
\label{VAR}
\end{figure}

Next, we used normalized mutual information (NMI) \cite{danon} to study similarities in partitions within and between community detection algorithms and between the full and super node representations. With the Louvain algorithm and with fitting SBMs according to the inference described in Ref. \cite{tiagosbm}, sampling and non-convexity in the optimization landscape lead to large variability in partitions. In Figure \ref{VAR} we computed normalized mutual information (NMI) between all pairs of partitions within and between multiple runs of Louvain and SBM fitting. A high normalized mutual information between a pair of partitions indicates that the algorithms identified similar community structures. The distribution of these pairwise NMI values for the full network and super node representation are visualized with box plots in Figure \ref{VAR}A and B, respectively. For each network, the purple, orange and maroon boxes show the distribution of pairwise NMIs between 10 runs of the Louvain, SBM fitting, and all individual runs of Louvain against all runs of SBM fitting, respectively.  The most significant improvement we observe under the super node representation is between runs of the Louvain algorithm and SBM fitting, suggesting that the new compressed representation of the network has prominent structural features that are robustly identified with both approaches. The Louvain algorithm is generally less variable than fitting an SBM, but we also observed decreased variability in the fitting of stochastic block models on the super node representation. 

While we see a reduction in variability in partitions on the super node representation, we also quantified the similarity in partitions on the full network under default parameters to those on the super node network by computing the NMI between their respective partitions, NMI$({\bf z}^{Full},{\bf z}^{SN})$. In Figure \ref{VAR}, we indicated the mean pairwise NMI between a full network and 5 super node partitions (again with $S=500$ super nodes) with a star. That is, a star in the Louvain (SBM) column indicates the mean pairwise NMI computed from NMI$({\bf z}^{Full},{\bf z}^{SN})$ using Louvain (SBM). As previously mentioned, a challenge in objective evaluation of partition similarity between representations is the difference in community size distributions. As a result, it is quite difficult to achieve a high NMI using default parameters. However, even these NMIs indicate structural similarity between partitions. Given the high variability within and between community detection algorithms indicated by moderate-to-low NMI values, one does not lose significantly more information by simply using the compressed super node network representation, even without matching scales. While these experiments are performed with $S=500$, we study the effects from using other $S$ values in supplementary tables 1 and 2. 

\begin{figure}
\centering
\includegraphics[scale=.4]{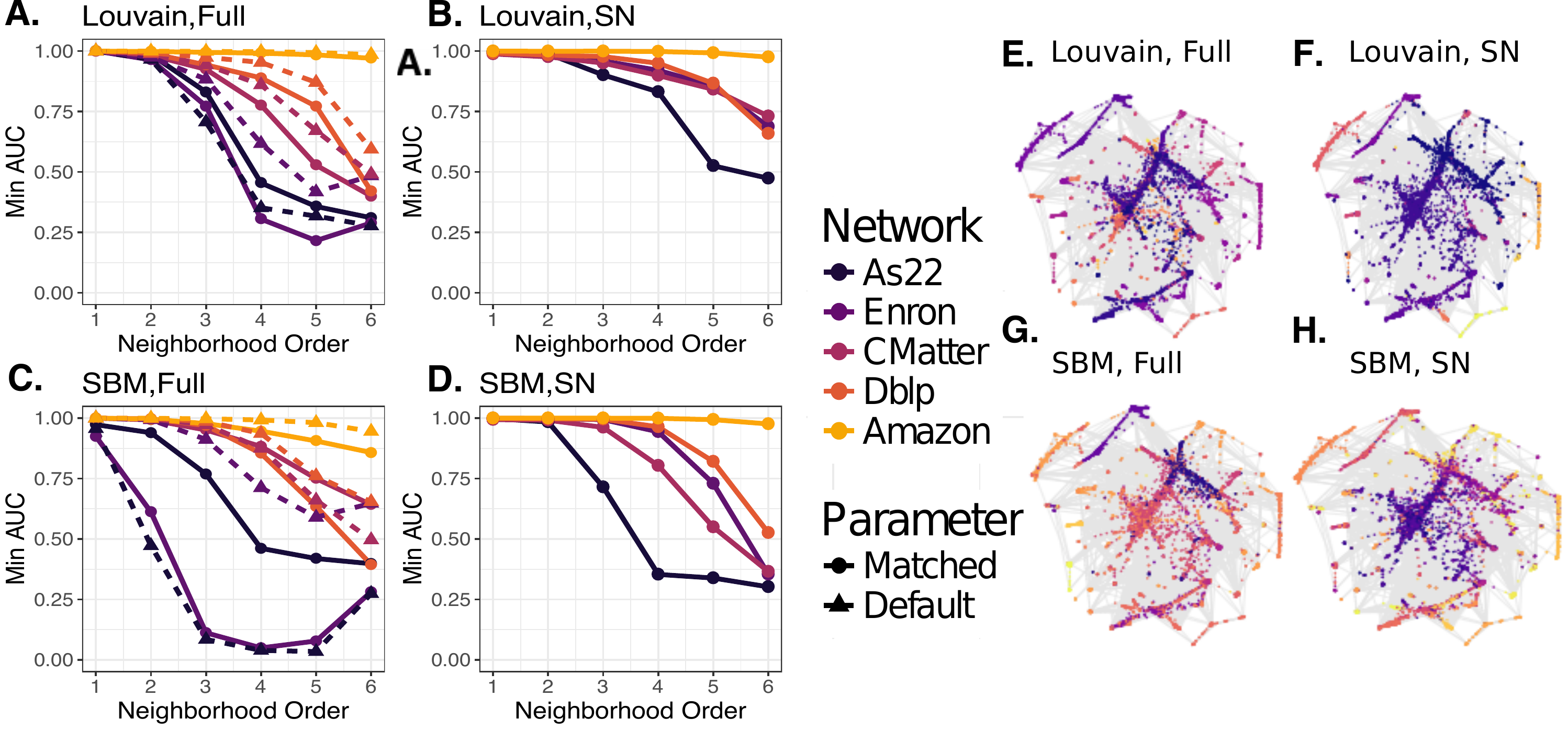}
\caption{{\bf Agreement of community assignments with local connectivity}. We study how  consistent partitions are within local neighborhood regions of the network by examining how well a node's neighbors (for various order neighborhoods) can be used to predict its community assignment, under some community partition ${\bf z}$. For each community in a partition, we give a binary prediction of whether a node is assigned to that community, based on probabilities we compute for a node from its neighbors. Sweeping the parameter $p$ that sets the probability required for a node to be assigned to a community, we compute ROC curves for each community and report the minimum AUC value observed. Panels {\bf A-D} show minimum AUC values observed as a function of neighborhood order for communities obtained from the full networks and super node representations by Louvain and by SBM. Line color indicates network and line type indicates communities obtained from the matched and default parameters used by the algorithms on the full networks. Panels {\bf E-H} visualize the communities obtained in the As22 data on the full network (default parameters) and super node representation (SN) under Louvain and SBM, with node colors indicating community memberships. }
\label{Viz}
\end{figure}

While we have emphasized benefits in the mechanics and usability of running standard community detection algorithms, we now seek to address whether the communities that we find using the super node representation align with local network connectivity so that neighbors are more likely to have similar community assignments and how this alignment compares with what we would have found by community detection on the full network. While we visualize this qualitatively for the As22 network in Figure \ref{Viz}E-H, we also designed a prediction task to quantify this alignment. In this prediction task, we seek to take a node-to-community partition (from either the full or super node network representations ${\bf z}^{Full}$ or ${\bf z}^{SN}$, respectively) and the full network ${\mathcal X}$ to see how accurately we can predict members of a community for different neighborhood sizes.
For network ${\mathcal X}$ with node-to-community assignments ${\bf z}$, we assign a probability distribution to each node over all of the communities under ${\bf z}$. For a neighborhood order $o$ (x-axis in Figure \ref{Viz}), we say that node $i$ has probability of being in community $k$, based on what fraction of its neighbors belong to that community under ${\bf z}$. Then for each community in ${\bf z}$, we perform a binary prediction task for whether each node of ${\mathcal X}$ should be assigned to that community, according to the computed probability distributions for all nodes with respect to that community. 

We sweep the probability parameter, $p$, representing the required threshold probability for a node to achieve in order to be assigned to a community in this binary classification task. By sweeping $p$ for each of the communities, we compute an ROC curve for each community and the corresponding areas under the curve (AUC). Finally, we use the minimum AUC value as our summary statistic of this task, with a high AUC value indicating that the neighboring regions of a node were strong predictors of community assignments, as shown in Figure \ref{Viz}A-D. All experiments are performed on 5 networks (As22, Enron, CMatter, Dblp, Amazon) and for both the matched and default parameters (indicated by line type) in the full network. (Recall from section 3.1 that the matched parameters for the full network were chosen based on the super node partition results under default settings; hence, there is no corresponding `matched' set for the super nodes in these plots.) We observe in most cases using the super node representation improves the minimum AUC value, indicating that communities obtained from this representation have higher agreement with local connectivity by this measure.
To qualitatively evaluate how the super node representation is able to ultimately partition the network into communities that are locally relevant, we visualize the As22 network, with nodes colored by communities identified under the full network (E,G) and super node representation (F,H) under Louvain (E,F) and SBM (G,H). Consistent with the quantitative results in Figure \ref{Viz}A-D, we observe that the super node representation leads to an effective coarse-graining, with the community labels appearing to be qualitatively consistent across large regions of the network.

\section{Conclusion and Future Work}
\label{sec:conclusion}

We developed an approach for compressing a network into a super node representation that can be used in standard community detection algorithms. Using the smaller super node network reduces runtime and the variability between multiple runs of the same community detection algorithm. Our results also demonstrate that the communities in the super node network are better aligned with local network neighborhoods in a predictive sense, while still being in relatively good alignment with the partitions obtained using the full network.

Super nodes may be useful in a variety of contexts where large datasets are otherwise difficult to mine and interpret. For example, one might visualize the super node version of the network rather than the entire network, or use the members of a super node to identify redundant information in the network. Future work on super node representations could include the extension of this method to directed, signed, attributed, or bipartite networks. Additionally, one might consider a probabilistic model framework, attempting to infer latent super node assignments. 

\clearpage{}

%

\bibliographystyle{unsrt}
\bibliography{NipsBib.bib}

\begin{thebibliography}{10}

\bibitem{fortunato2010}
Santo Fortunato.
\newblock Community detection in graphs.
\newblock {\em Physics Reports}, 486(3):75--174, 2010.

\bibitem{fortunato}
Santo Fortunato and Darko Hric.
\newblock Community detection in networks: A user guide.
\newblock {\em Physics Reports}, 659:1--44, 2016.

\bibitem{portercommunity}
Mason~A Porter, Jukka-Pekka Onnela, and Peter~J Mucha.
\newblock Communities in networks.
\newblock {\em Notices of the AMS}, 56(9):1082--1097, 2009.

\bibitem{sbmOrig}
Tom~AB Snijders and Krzysztof Nowicki.
\newblock Estimation and prediction for stochastic blockmodels for graphs with
  latent block structure.
\newblock {\em Journal of classification}, 14(1):75--100, 1997.

\bibitem{bigclam}
Jaewon Yang and Jure Leskovec.
\newblock Overlapping community detection at scale: a nonnegative matrix
  factorization approach.
\newblock In {\em Proceedings of the sixth ACM international conference on Web
  search and data mining}, pages 587--596. ACM, 2013.

\bibitem{tiagosbm}
Tiago~P Peixoto.
\newblock Efficient {Monte Carlo} and greedy heuristic for the inference of
  stochastic block models.
\newblock {\em Physical Review E}, 89(1):012804, 2014.

\bibitem{louvain}
Vincent~D Blondel, Jean-Loup Guillaume, Renaud Lambiotte, and Etienne Lefebvre.
\newblock Fast unfolding of communities in large networks.
\newblock {\em Journal of statistical mechanics: theory and experiment},
  2008(10):P10008, 2008.

\bibitem{newmanmodularity}
Mark~EJ Newman.
\newblock Modularity and community structure in networks.
\newblock {\em Proceedings of the national academy of sciences},
  103(23):8577--8582, 2006.

\bibitem{gleich}
David~F Gleich and C~Seshadhri.
\newblock Vertex neighborhoods, low conductance cuts, and good seeds for local
  community methods.
\newblock In {\em Proceedings of the 18th ACM SIGKDD international conference
  on Knowledge discovery and data mining}, pages 597--605. ACM, 2012.

\bibitem{spectralCommunity}
Mark~EJ Newman.
\newblock Spectral methods for community detection and graph partitioning.
\newblock {\em Physical Review E}, 88(4):042822, 2013.

\bibitem{browet}
Arnaud Browet, P-A Absil, and Paul Van~Dooren.
\newblock Community detection for hierarchical image segmentation.
\newblock In {\em International Workshop on Combinatorial Image Analysis},
  pages 358--371. Springer, 2011.

\bibitem{slic}
Radhakrishna Achanta, Appu Shaji, Kevin Smith, Aurelien Lucchi, Pascal Fua, and
  Sabine S{\"u}sstrunk.
\newblock Slic superpixels compared to state-of-the-art superpixel methods.
\newblock {\em IEEE transactions on pattern analysis and machine intelligence},
  34(11):2274--2282, 2012.

\bibitem{jureTruth}
Jaewon Yang and Jure Leskovec.
\newblock Defining and evaluating network communities based on ground-truth.
\newblock {\em Knowledge and Information Systems}, 42(1):181--213, 2015.

\bibitem{gilbert}
Anna~C Gilbert and Kirill Levchenko.
\newblock Compressing network graphs.
\newblock In {\em Proceedings of the LinkKDD workshop at the 10th ACM
  Conference on KDD}, volume 124, 2004.

\bibitem{supergenomic}
Andreas~Martin Lisewski, Joel~P Quiros, Caroline~L Ng, Anbu~Karani Adikesavan,
  Kazutoyo Miura, Nagireddy Putluri, Richard~T Eastman, Daniel Scanfeld, Sam~J
  Regenbogen, Lindsey Altenhofen, et~al.
\newblock Supergenomic network compression and the discovery of exp1 as a
  glutathione transferase inhibited by artesunate.
\newblock {\em Cell}, 158(4):916--928, 2014.

\bibitem{SuperNodeSide}
Liang Yang, Di~Jin, Dongxiao He, Huazhu Fu, Xiaochun Cao, and Francoise
  Fogelman-Soulie.
\newblock Improving the efficiency and effectiveness of community detection via
  prior-induced equivalent super-network.
\newblock {\em Scientific Reports}, 7(1):634, 2017.

\bibitem{CoreHD}
Lenka Zdeborov{\'a}, Pan Zhang, and Hai-Jun Zhou.
\newblock Fast and simple decycling and dismantling of networks.
\newblock {\em Scientific Reports}, 6, 2016.

\bibitem{jungirie}
Kyomin Jung, Wooram Heo, and Wei Chen.
\newblock Irie: Scalable and robust influence maximization in social networks.
\newblock In {\em Data Mining (ICDM), 2012 IEEE 12th International Conference
  on}, pages 918--923. IEEE, 2012.

\bibitem{kempe}
David Kempe, Jon Kleinberg, and {\'E}va Tardos.
\newblock Maximizing the spread of influence through a social network.
\newblock In {\em Proceedings of the ninth ACM SIGKDD international conference
  on Knowledge discovery and data mining}, pages 137--146. ACM, 2003.

\bibitem{danon}
Leon Danon, Albert Diaz-Guilera, Jordi Duch, and Alex Arenas.
\newblock Comparing community structure identification.
\newblock {\em Journal of Statistical Mechanics: Theory and Experiment},
  2005(09):P09008, 2005.

\end{thebibliography}



%
%
%

\end{document}